\documentclass[lettersize,journal]{IEEEtran}
\usepackage{amsmath,amsfonts}
\usepackage{algorithmic}
\usepackage{algorithm}
\usepackage{array}
\usepackage[caption=false,font=normalsize,labelfont=sf,textfont=sf]{subfig}
\usepackage{textcomp}
\usepackage{stfloats}
\usepackage{url}
\usepackage{verbatim}
\usepackage{graphicx}
\usepackage{cite}
\usepackage{physics}

\usepackage{tikz}

\usepackage{graphicx} 

\usepackage{xcolor}

\usepackage{acronym}
\newacro{BSM}[BSM]{Bell state measurement}
\newacro{BSMs}[BSM]{Bell state measurements}

\newacro{LLEG}[LLEG]{link-level entanglement generation}
\newacro{AQM}[AQM]{Active Queue Management}
\newacro{RED}[RED]{Random Early Detection}
\newacro{PI}[PI]{Proportional Integral}
\newacro{ECN}[ECN]{Explicit Congestion Notification}

\newacro{TCP}[TCP]{Transmission Control Protocol}
\newacro{RAP}[RAP]{Rate Adaptation Protocol}
\newacro{QTCP}[QTCP]{Quantum TCP}
\newacro{IP}[IP]{Internet Protocol}
\newacro{e2e}[e2e]{end-to-end}
\newacro{FIFO}[FIFO]{First In First Out}

\newacro{QKD}[QKD]{quantum key distribution}

\newacro{AIMD}[AIMD]{Additive Increase, Multiplicative Decrease}

\newacro{OSPF}[OSPF]{Open Shortest Path First}
\newacro{RTT}[RTT]{Round Trip Time}

\newcommand{\REVISION}[1]{{{#1}}}

\providecommand{\myparab}[1]

\newcommand{\eat}[1]{}

\title{Leveraging Internet Principles to Build a Quantum Network}

\author{Leonardo~Bacciottini,
Matheus~Guedes~De~Andrade,
Shahrooz~Pouryousef,
Emily~A.~Van~Milligen,
Aparimit~Chandra,
Nitish~K.~Panigrahy,
Nageswara~S.~V.~Rao,
Gayane~Vardoyan,
Don~Towsley

\thanks{L.B., M.G.D.A, S.P., A.C., G.V., and D.T. are with University of Massachusetts Amherst, USA}%
\thanks{E.A.V.M. is with The University of Arizona, USA}%
\thanks{N.K.P. is with SUNY Binghamton, USA}%
\thanks{N.S.V.R. is with Computational Sciences and Engineering Division, Oak Ridge National Laboratory, USA
}
\thanks{When this research work started, L. B. was with University of Florence and Dept. of Information Engineering (DII), University of Pisa, ITALY}
}

\begin{document}
\date{}
\maketitle

\begin{tikzpicture}[remember picture,overlay] \node[anchor=south,yshift=0pt] at (current page.south) {\fbox{\parbox{\dimexpr\textwidth-\fboxsep-\fboxrule\relax}{\tiny
    © 2024 IEEE.  Personal use of this material is permitted.  Permission from IEEE must be obtained for all other uses, in any current or future media, including reprinting/republishing this material for advertising or promotional purposes, creating new collective works, for resale or redistribution to servers or lists, or reuse of any copyrighted component of this work in other works.
  }
}};
\end{tikzpicture}

\begin{abstract}
Designing an operational architecture for the Quantum Internet is challenging in light of both fundamental limits imposed by physics laws and technological constraints. Here, we propose a method to abstract away most of the quantum-specific elements and formulate a best-effort quantum network architecture based on packet switching, akin to that of the classical Internet. \REVISION{This} reframing provides an opportunity to exploit the \REVISION{many available and well-understood protocols} within the Internet context. As an illustration, we tailor and adapt classical congestion control and active queue management protocols to quantum networks, employing an architecture wherein quantum end and intermediate nodes effectively regulate demand and resource utilization, respectively. Results show that these classical networking tools can be effective in managing quantum memory decoherence and maintaining end-to-end fidelity around a target value.
\end{abstract}

\section{Introduction}







\IEEEPARstart{Q}{uantum} networks enable
 interactions between remote quantum systems, thereby realizing distributed quantum applications with unique capabilities beyond purely classical techniques.
Once realized, a full-fledged Quantum Internet will work in concert with the classical Internet to support 
unprecedented services, including, \ac{QKD} \cite{Munro_2015}, distributed, cloud, and blind quantum computing \cite{VanMeter_2016}, clock synchronization \cite{Degen_2017}, and quantum-enhanced sensing \cite{Degen_2017}. 
However, quantum networks face a number of technological and fundamental challenges, in part due to the no-cloning theorem \cite{Munro_2015}, which precludes the use of classical signal amplifiers and routers for long-distance quantum state transport.
Quantum entanglement distribution (i.e., two-way) networks provide near-term solutions using quantum repeaters and switches, which can outperform advanced one-way solutions based on quantum error correction \cite{Mantri2024}.

Quantum entanglement is a stronger-than-classical correlation between quantum states. When Alice and Bob share a maximally entangled qubit pair, i.e., Bell pair, the quantum teleportation protocol enables the transfer of an arbitrary single-qubit state from Alice to Bob.  Teleportation consists of a \ac{BSM}---after which the shared Bell pair is destroyed---followed by the transmission of two classical bits of information to enable Bob to recover the original state.

Teleportation can also be used \REVISION{to generate and distribute} entanglement between Alice and Bob. If Charlie holds two qubits, the first entangled with a qubit at Alice and the second entangled with a qubit at Bob, then he can teleport the first qubit to Bob. The outcome is a single Bell pair shared directly between Alice and Bob. Thus, Charlie acted as a \emph{quantum repeater} by performing \emph{entanglement swapping} between two separate Bell pairs. In quantum networks, Bell pairs are first generated and shared between physically connected quantum \REVISION{nodes}, a process called \ac{LLEG}, and then the pairs are stitched together via swapping to realize end-to-end entanglement. 

Entanglement swapping can in principle be iterated to reach arbitrarily long distances, but quantum noise imposes practical limitations on the maximum number of consecutive swapping operations. Swapping imperfect Bell pairs exponentially degrades the fidelity (a measure of quality) of entanglement, even \REVISION{under ideal} operations. Moreover, quantum storage introduces additional decoherence that depends on the time the \REVISION{Bell pairs spend} in storage. These limiting noise factors \REVISION{must be taken into account} in the design of an operational quantum network.

At first glance, it may seem that quantum and classical networks are conceptually different. Over the last few years, several works tried to partially bridge the gap between the two domains. Among these, Dahlberg \textit{et al.} \cite{Dahlberg_2019} were the first to introduce a quantum protocol stack that mirrors in shape the well-known Transmission Control Protocol/Internet Protocol (TCP/IP) suite. Van Meter \textit{et al.} \cite{van2022quantum} designed a recursive architecture incorporating elements from the Software-Defined Networking paradigm. Zhao \textit{et al.} \cite{zhao2023distributed} started a line of research where they design quantum memory allocation algorithms inspired by TCP's congestion control. Previous works investigated packet-switched or connectionless quantum networks \cite{Shabani_2022, Li_2023_Entanglement, Yooetal2024, Xiao_2024}, but they \REVISION{employ only optical} switching \cite{Shabani_2022}, require reserved paths \cite{Li_2023_Entanglement} or circuits \cite{Yooetal2024}, or are limited to the Network layer \cite{Xiao_2024}.

This work showcases that classical networking principles can be applied to quantum networks by designing a two-way quantum network architecture modeled after a classical packet-switched network.
We achieve this result by (i) mapping the entanglement swapping sequence to hop-by-hop forwarding of a \emph{quantum datagram} packet, and (ii) representing the quantum memories of devices as queues of limited size for these datagrams. At a higher level, we introduce a Transport layer protocol that orchestrates entanglement requests into \ac{e2e} flows and uses \ac{AIMD} congestion control mechanisms and \ac{AQM}, borrowed from classical networking, to enforce fair resource sharing.

Simulations show that this architecture provides the flexibility to control e2e fidelity around a target value, whose range depends on the underlying hardware characteristics. However, \REVISION{increasing} e2e fidelity inevitably lowers aggregate throughput, highlighting a fundamental trade-off between these two competing \REVISION{performance measures}.


The contributions of this work are as follows:
\begin{itemize}
    \item We propose a connectionless, best-effort architecture for quantum networks;
    \item To illustrate how this architecture allows the application of classical networking tools, we adapt classical congestion control policies to quantum networks;
    \item We demonstrate how this architecture counters quantum-memory decoherence through simulated experiments.
\end{itemize}




\section{Architecture Overview}\label{sec:sys_model}

    Entanglement distribution networks enable the sharing of entangled states among remote network nodes. In this section, we introduce the building blocks of a packet-switched entanglement distribution network.

    \subsection{Architectural Components}

        \begin{figure}
        \centering
        \includegraphics[width=\linewidth]{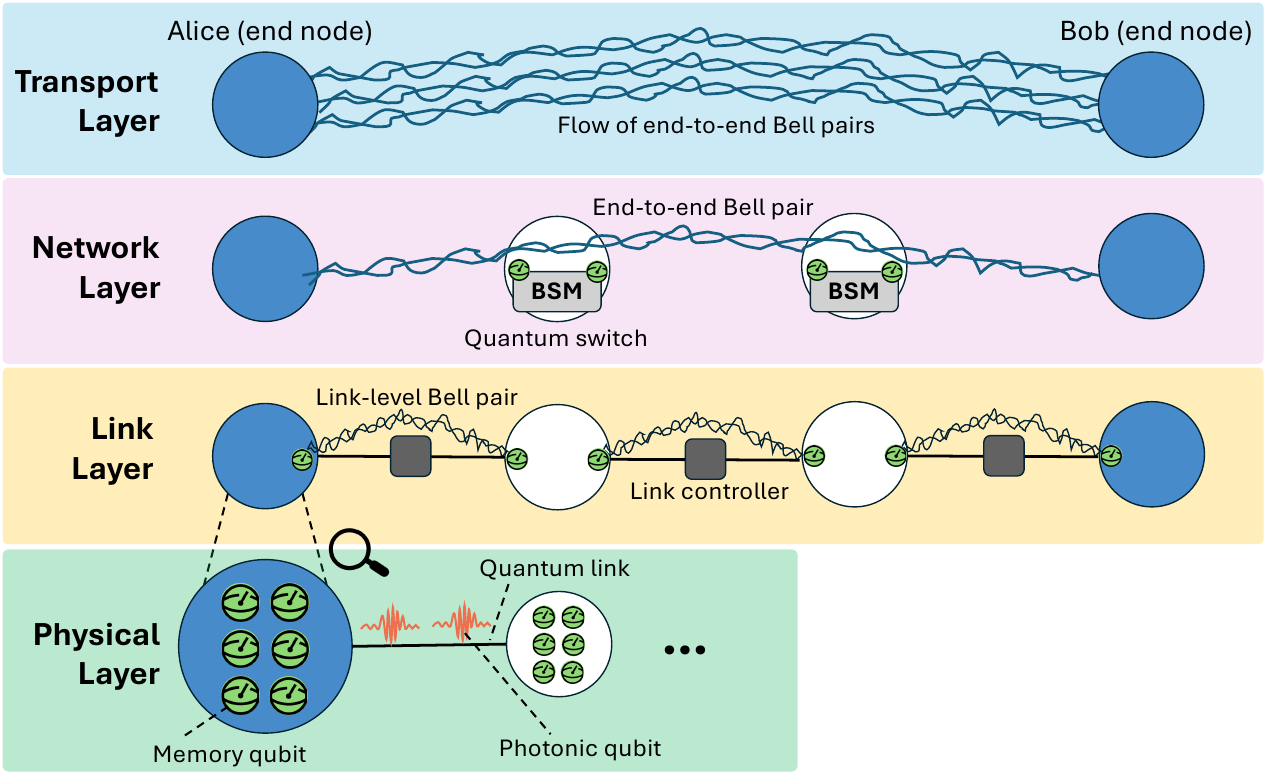}
        \caption{
        A functional view of the quantum network architecture and its components, organized according to the adopted protocol stack.}
        \label{fig:sys_model}
        \end{figure}

    Quantum network nodes are processors equipped with quantum memories and circuitry to implement quantum operations. At a high level, quantum networks interconnect nodes through quantum links capable of generating entanglement. The quantum network architecture proposed in this work builds upon the following components and elementary operations.

        \subsubsection{Quantum Links}
             are the physical media---e.g., fiber or free space---to propagate photons. They enable entanglement generation between two neighboring nodes. Every quantum link is managed by a dedicated \emph{link controller } that multiplexes the quantum link across multiple competing users. Nodes connected by a quantum link are called \emph{adjacent} nodes.

        \subsubsection{End Nodes}
            also referred to as \emph{users}, end nodes generate \emph{entanglement requests} and must be attached to a quantum link.

        \subsubsection{Quantum Switches}
             are the infrastructure nodes responsible for carrying out \emph{entanglement swapping} to distribute entanglement to the end nodes.
            A quantum switch has several interfaces, each attached to a quantum link. Generally, a switch \REVISION{is also a router, as it makes} decisions about entanglement swapping. We assume all quantum switches and end nodes can classically communicate with each other.

        \subsubsection{Quantum Memories}
            each node possesses a number of memory qubits capable of storing quantum states. \REVISION{We indicate} \emph{coherence time} of a memory qubit \REVISION{as} the storage time until the error probability reaches $1 - e^{-1}$. The \REVISION{effect} of the error depend on the chosen noise model. For example, depolarizing noise transforms the stored state to a completely mixed state.
            
            
            
            In contrast with quantum memories, photonic qubits, also known as flying qubits, are qubit states encoded into light pulses transmitted through quantum links. 
            
            

        \subsubsection{Bell State Measurements (BSM)}
            measurement of a quantum system collapses its state as specified by its outcome. A crucial kind of measurement in quantum information is the BSM, which produces two classical bits of information by projecting the state of two qubits onto the Bell orthonormal basis. The entanglement swapping protocol employs a \ac{BSM} operation on two qubits held by a quantum \REVISION{switch}. To \REVISION{determine} which of the four Bell states is obtained after the procedure, the two-bit outcome is transmitted to \REVISION{one} of the two nodes, \REVISION{where} local rotations are performed. When entanglement swapping is iterated more than once, it suffices to keep track of the bitwise sum of all the \ac{BSM} operations.
            \REVISION{Our} architecture design supports both deterministic and probabilistic BSM implementations.

    \subsection{Protocol Stack}
        
        Protocol stacks are used to \REVISION{simplify network models by breaking them into layered protocols}, where each layer \REVISION{provides} a service to the layer above by \REVISION{utilizing} the service of the layer below.
        

        We adopt a quantum network protocol stack mirroring the five layers of the classical Internet's TCP/IP suite. It consists of the following  layers:
        
        \begin{enumerate}
            \item The \emph{Physical} layer encompasses quantum hardware and the necessary classical control hardware to implement quantum operations in the nodes, such as entanglement swapping protocols, 
            operations on memory qubits, and intermediate steps utilized in entanglement generation. 
            \item The \emph{Link} layer \REVISION{provides an interface for requesting and obtaining heralded} link-level entanglement.
            \item The \emph{Network} layer consumes link-level pairs and generates \ac{e2e} entanglement through entanglement swapping. \REVISION{Like} its classical counterpart, it is responsible for path selection, node discovery, addressing, and routing.
            \item The \emph{Transport} layer manages flows of Bell pairs from an \ac{e2e} perspective.
            \item Finally, the \emph{Application} layer implements custom logic on top of the Transport layer service. Every application session instantiates at least one \ac{e2e} flow of Bell pairs.
        \end{enumerate}
        
        Distillation protocols \cite{Munro_2015} probabilistically produce higher-fidelity Bell pairs. Although not specifically included in this stack, they can be introduced at the Link layer, Network layer, and/or, e2e (Transport or Application layer).
        
        We summarize in Fig. \ref{fig:sys_model} the architectural elements introduced in this section and their placement in the protocol stack. 
        

    \section{Link Layer} \label{sec:link}

        The Link layer abstracts away the physical platform and control hardware of a quantum link. It \REVISION{provides} an interface to request and (eventually) obtain Bell pairs shared among the two nodes on either side of a quantum link.
        
        The functionality of the Link layer is realized by \REVISION{the LLEG process} and link controllers: respectively a southbound and a northbound component with respect to the interactions \REVISION{between Link and} neighboring layers (Physical and Network).

        \subsection{Link-level Entanglement Generation (LLEG)}
            \ac{LLEG} is the elementary process that distributes (imperfect) Bell pairs to two \REVISION{nodes} connected by a quantum link. The direct connection allows photons to be physically transmitted between the two \REVISION{nodes}, enabling generation of entanglement between memory qubits interfacing with the quantum link. We refer to Bell pairs generated by \ac{LLEG} as \emph{link-level pairs}. 
            
            Several \ac{LLEG} protocols exist \cite{Munro_2015}, some of which depend on the adopted hardware platform.
            For example, the entangled photon source can be placed midpoint or at either edge of the quantum link.
            What is common to all these protocols is that \ac{LLEG} is intrinsically probabilistic: attempts succeed with a probability $p$ that depends on factors such as \REVISION{optical BSMs}, the efficiency and brightness of the photon source, the efficiency of frequency coupling, medium attenuation, and link length. 
            While the proposed architecture does not require a specific configuration, \REVISION{for illustrative purposes we adopt midpoint source placement in this work.}

        \subsection{Link Controllers}
            
            Link controllers determine when \ac{LLEG} on a specific quantum link is activated, and serve as the interface between the Link and Network layers.
            
            Nodes can request a link-level pair at any time from any quantum link to which they are attached. Every request is associated with a unique sequence number to differentiate between multiple requests. When the pair is generated, the link controller schedules a pending request to fulfill and returns two pieces of information: (i) a \emph{link-level label}, which identifies the new pair, and (ii) the sequence number of the served request. Sequence numbers are used to multiplex requests from different nodes, while link-level labels function as an addressing scheme for link-level pairs between adjacent nodes. Unlike MAC addresses \cite{Kurose_2012} in the classical Internet, these labels reference the memory qubits holding the link-level pair rather than the nodes themselves.
            
            The architecture is agnostic of the scheduling policy on pending requests. For example, a simple policy adopted in our experiments is to pick the youngest request from the node with the largest backlog. Link controllers may also decide whether to activate \ac{LLEG} only when there is at least one pending request (\emph{on-demand} approach) or to keep it always active and cache link-level pairs if possible (\emph{continuous} approach).
            
            Depending on the \ac{LLEG} protocol used, it is convenient to place the link controller either at the midpoint of the link or at one of the two connected nodes to reduce delays introduced by the propagation of classical information required by the controllers.
    
    \section{Network Layer} \label{sec:network}
    
       The task of the Network layer is to \REVISION{create \ac{e2e} entanglement from link-level pairs} by finding a path that connects two end nodes, and carrying out entanglement swapping at intermediate switches.
       
       The Network layer mirrors three key properties of its classical counterpart IP: it is (i) packet-switched, as every Bell pair is treated and forwarded on the network like a packet. It is (ii) connectionless, meaning that each packet is handled independently without reserving network resources in advance (stochastic multiplexing). Additionally, (iii) it operates on a best-effort basis: the Network layer does not guarantee reliability. Bell pairs may be discarded due to noise, errors, or congestion. Any \REVISION{\ac{e2e}} quality-of-service features, such as entanglement regeneration and flow rate control, must be implemented at the Transport or Application layers. Other multiplexing methods at the Network layer---suitable for resource-reserving solutions---include time-slotted circuit switching and quantum memory partitioning.
        
       We assume that every node obtains a unique address in the quantum network in the same fashion as with IP addressing.
       

       \subsection{Sequential Swapping}

\begin{figure}
                \centering
                    \includegraphics[width=.8\linewidth]{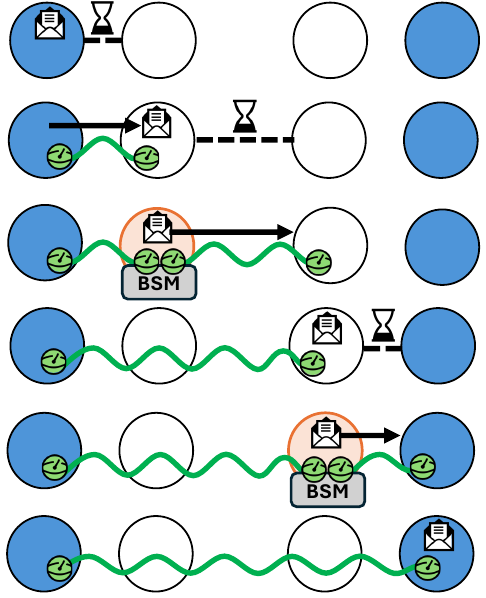}
                  \caption{Sequential entanglement swapping. Hourglasses represent the time elapsed while the node waits for the next link-level pair. Sequential swapping is implemented with the help of classical packets (shown as envelopes) forwarded hop by hop as \acp{BSM} occur. As each packet travels along the path, it solicits the next node and collects the \ac{BSM} outcomes required by the destination node to determine the \ac{e2e} Bell state. \REVISION{The classical packet and the moving half of the Bell pair constitute together a quantum datagram}. \label{fig:timing_seq}
                  }
                \end{figure}

            \REVISION{We} map entanglement swapping into a hop-by-hop forwarding operation. We achieve this by adopting a \emph{sequential entanglement swapping policy}, showcased in Fig.~\ref{fig:timing_seq}.
            
            \REVISION{Suppose that} an end node wants to share a Bell pair with another end node. An intermediate switch on the path connecting them requests a link-level pair on the next link only after it is solicited through a classical signal from the previous node. \REVISION{Once} the next link-level pair is ready, the switch \REVISION{performs} a \ac{BSM} and solicits the next hop, also attaching the \ac{BSM} outcome. A similar sequential approach at the Network layer has also been recently adopted in \cite{Xiao_2024}.

        \subsection{Switching Quantum Datagrams}\label{sebsec:switching}

             The key component of the sequential swapping protocol is the quantum datagram, or \textit{q-datagram}. A q-datagram is a packet that identifies a Bell pair shared between an end node and a quantum switch. \REVISION{It contains} both classical and quantum data about the Bell pair it represents. The quantum data is a qubit holding \REVISION{one} half of the Bell pair, while the classical data includes meta-information such as:
            \begin{itemize}
                \item source node address;
                \item destination node address;
                \item sequence number, used when requesting the next link-level pair;
                \item flow identifier, used by the Transport layer to map the Bell pair to an \ac{e2e} flow;
                \item link-level label, used to map the q-datagram to a memory qubit (updated at every hop);
                \item Pauli frame correction, i.e., the amalgamation of all \ac{BSM} outcomes so that the destination can decode the final Bell state (updated at every hop).
            \end{itemize}

            As they forward q-datagrams, quantum switches teleport the associated qubit state, pushing it toward the destination end node.
            
            When a quantum switch receives the classical portion of a q-datagram, it knows that one of its memory qubits -- retrievable via the link-level label -- is entangled with the source node. The switch then:
            \begin{enumerate}
                \item Determines the next hop toward the destination end node;
                \item Requests a link-level Bell pair from the link controller on the next quantum link, specifying the sequence number of the q-datagram;
                \item Carries out a \ac{BSM} between the q-datagram qubit and the qubit from the link-level Bell pair when the pair is generated;
                \item Updates the link-level label and the Pauli frame correction fields of the q-datagram;
                \item Forwards the classical contents of the q-datagram to the next hop.
            \end{enumerate}

            There can be a delay between steps 2 and 3. Such a delay depends on Link and Physical layer features such as quantum link attenuation, entanglement attempt rate, and link controller scheduling policy. Quantum switches, therefore, may have to buffer the quantum and classical components of q-datagrams. This requires the deployment of quantum memory and classical memory management algorithms. Responsibilities of \REVISION{a \emph{quantum memory management unit} inside a node} include allocation of memory to newly arriving q-datagrams, decision as to which q-datagram to remove when the quantum memory is full, and possibly implementing storage cutoffs, i.e. discarding an overly degraded quantum state.

            The Network layer protocol exploits the sequence number mechanism of link controllers to determine which q-datagram it can forward (steps \REVISION{3, 4, 5} above) among the pool of buffered q-datagrams.

            \vspace{.2cm}
            \textbf{Takeaway:} \textit{The Network layer treats Bell pairs as quantum datagrams and maps entanglement swapping into datagram forwarding, thereby \REVISION{representing the quantum network as a packet-switched system}.} 
            \vspace{.2cm}

        \subsection{Considerations}

            The proposed Network layer requires a routing protocol for nodes to determine the next hop of a q-datagram. Fortunately, this architecture supports classical network routing algorithms, such as \ac{OSPF} \cite{Kurose_2012}. Alternatively, quantum-specific routing protocols can be considered.

            Additionally, there is potential to include supplementary metadata in q-datagrams, such as storage times at each node, quantum noise parameters of the traversed hops, or priority levels. These fields could provide valuable insights into the fidelity of the delivered Bell pairs, particularly when used alongside \ac{e2e} tomography. Furthermore, this metadata could enable the implementation of differentiated services for q-datagrams at the Network layer.

\section{Transport Layer}\label{sec:transport}


    Quantum applications will need more than one Bell pair. For example, \ac{QKD} protocols require batches of Bell pairs sized proportionally to the length of the shared keys. The Transport layer enables the creation and management of Bell pair \emph{flows} between end nodes. Building upon similarities between the quantum and classical Network layers, we propose a Transport layer protocol inspired by the classical TCP, which manages \ac{e2e} flows of q-datagrams similarly to how TCP manages flows of Internet packets. We refer to the proposed protocol as \ac{QTCP}.

    \ac{QTCP} borrows the \REVISION{port-based addressing scheme} from its classical counterpart and allows applications to instantiate flows. A \ac{QTCP} flow goes through three distinct phases: (i) an opening handshake, where the two end nodes agree on the flow creation, (ii) a transmission phase, where Bell pairs are generated, and (iii) a closing handshake, where the end nodes deallocate the flow after it has served its purpose. Note that quantum switches are not involved in any of these steps.

    In this paper, we focus on the transmission phase. During this phase, one of the two end nodes, known as the {\em flow source}, begins \REVISION{to generate} requests for Bell pairs between \REVISION{itself} and the other end node, \REVISION{known as} the {\em flow destination}. No assumption is made about how requests are generated: they can arrive as a single batch, one at a time, or in multiple batches depending on the application. When they enter the system, requests are translated into q-datagrams managed by the Network layer protocol. Upon delivery of a q-datagram to the flow destination, \ac{QTCP} returns an acknowledgment message to the flow source. \ac{QTCP} acknowledgments enable drop detection of q-datagrams, triggering their retrial and providing a reliable delivery service.

    If all requests are \REVISION{accepted} into the system without \REVISION{admission} control, then the quantum network can become severely congested. Congestion is detrimental to performance since it can increase buffering time, causing quantum decoherence with an exponential decay in the fidelity of stored entanglement. Also considering the network resources wasted due to frequent drops, congestion can completely disrupt the network service. With this motivation, we introduce the concept of a  \emph{quantum congestion controller}.
    Previous works tackle congestion in quantum networks by reserving quantum memories \cite{zhao2023distributed}, quantum link capacities \cite{Dahlberg_2019}, or usage time \cite{van2022quantum}. Instead, QTCP dynamically adapts the number of q-datagrams in the network.

    \subsection{\ac{QTCP} Congestion Control}

        A congestion controller uses an algorithm to determine when and how requests generated from the flow source are allowed into the quantum network.

        We can directly borrow any TCP congestion control algorithm (e.g., Reno, Vegas, Tahoe, or CUBIC \cite{Kurose_2012}) and apply it to \ac{QTCP}. In the following, we describe the algorithm we \REVISION{used} and the \REVISION{motivation} behind our design choices.
        We adopt an AIMD congestion control with congestion avoidance and slow start phases. We choose window-based congestion control, where the control knob (\emph{window size}) is the maximum number of q-datagrams that the flow can concurrently have in flight within the network. The congestion controller tracks the current number of q-datagrams in the network using QTCP acknowledgments.
        
        During the slow start phase, the algorithm increases the window size $w \gets w + 1$ every time the flow source receives a \ac{QTCP} acknowledgment. During the congestion avoidance phase, the window size is increased only after receiving $w$ acknowledgments (additive increase).
        
        Since the Network layer is a best-effort service, q-datagrams can be lost before they reach their destination. 
        The algorithm interprets q-datagram losses as congestion events. When a congestion event is detected, the algorithm halves the window size $w = w/2$ (multiplicative decrease).

        Many TCP versions rely on timeouts or out-of-order deliveries to detect losses. We instead propose an \ac{ECN} mechanism\footnote{Not to be confused with the ECN extension to the classical TCP/IP protocols \cite{Kurose_2012}.}, where the quantum switch that causes the loss immediately notifies the flow source through a dedicated classical message. This approach introduces additional overhead, but leads to an earlier reaction to congestion events that is crucial to \REVISION{curtailing} quantum memory decoherence.
        

    \subsection{Active Queue Management \REVISION{(AQM)}}\label{sec:aqm}

    It has been shown for the classical Internet that allowing packet queues to fill until they overflow, subsequently causing packet drops, leads to poor performance \cite{hollot_2001}. Since the proposed quantum network model shares similar characteristics, it is reasonable to expect similar consequences. One solution to this problem is to introduce AQM, which technically works at the Network layer, but it is so tightly related to congestion control that we place its description here for ease of exposition.

   Rather than allowing queues to fill and overflow, \ac{AQM} algorithms monitor queue lengths at routers and generate congestion signals to reduce traffic preemptively. \ac{AQM} algorithms reduce latency and packet loss while enhancing fairness.
    
    In the proposed architecture, every time a switch receives a q-datagram during time step $i$, it marks it as congested with a certain probability $p_i$ regulated by the \ac{AQM} algorithm. We assume that q-datagrams have an additional field to store this flag. The \ac{QTCP} acknowledgment message relays back the flag to the flow source, which interprets it as a congestion event. With a similar reasoning as with q-datagram loss, we can again use ECN instead of this mark-and-relay mechanism.

    Several \ac{AQM} algorithms have been developed, including \ac{RED} \cite{Kurose_2012} and \ac{PI} control \cite{hollot_2001}. We use \ac{PI} because it can regulate the queue around a \emph{target buffering time} for q-datagrams at each switch, making it easier to control memory decoherence. It is also fairly easy to tune according to the rules given in \cite{hollot_2001}. \ac{PI} control employs a control-theoretic approach, using proportional and integral components to periodically update the probability $p_i \rightarrow p_{i+1}$ between time steps $i$ and $i+1$ as follows \cite{hollot_2001}:
    \begin{equation}
        p_{i+1} = p_i + \alpha(t_{i+1} - t_{\text{\textit{tgt}}}) - \beta(t_{i} - t_{\text{\textit{tgt}}}),
    \end{equation}
    where $t_{\text{\textit{tgt}}}$ is the target buffering time, $t_i$ is the buffering time experienced at time step $i$, and $\alpha$, $\beta$ are tunable parameters with $\alpha>\beta>0$. We imagine independent \ac{PI} controllers installed on every quantum switch.
    


    \subsection{Putting it all together}
        \begin{figure*}
                \centering
                \includegraphics[width=\linewidth]{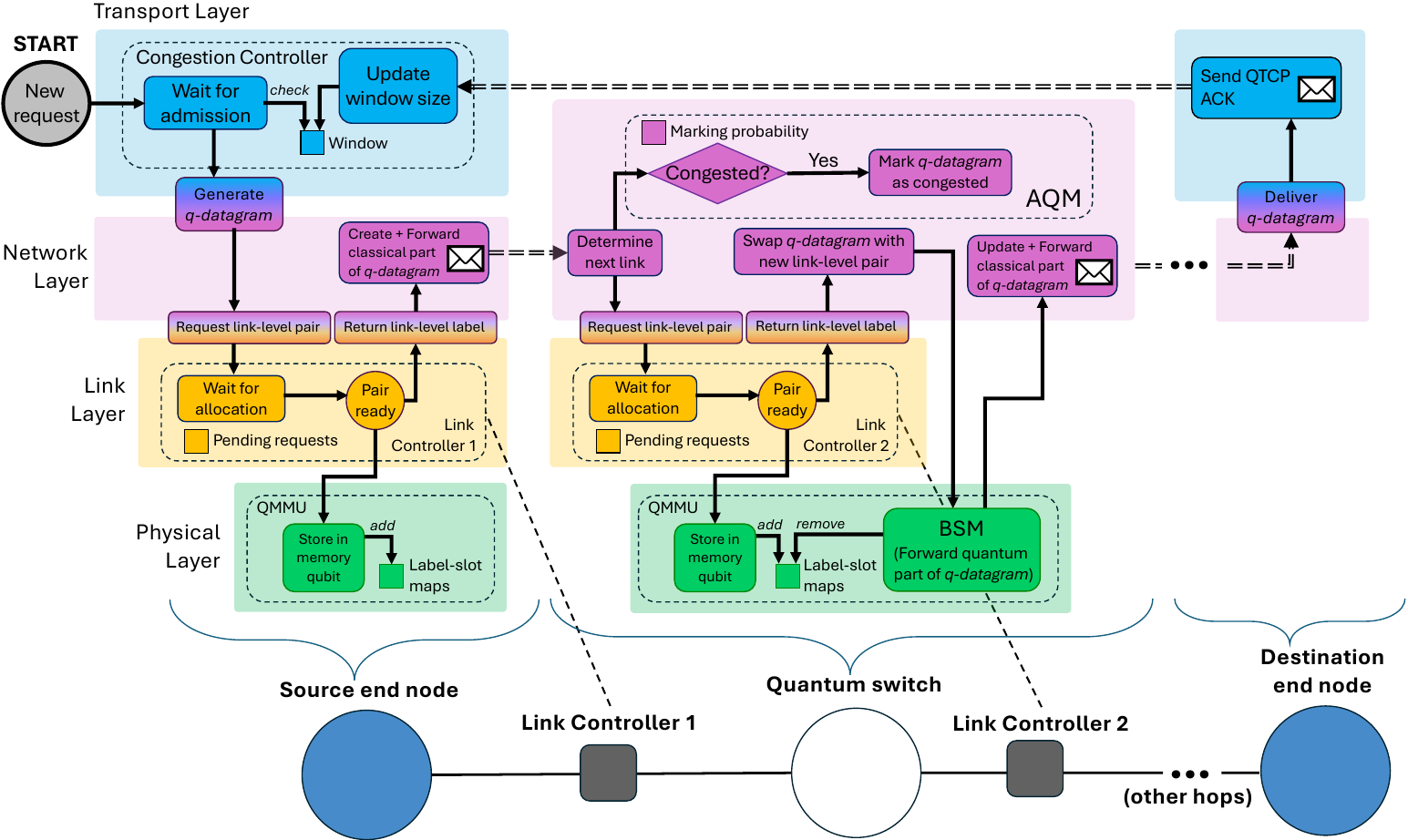}
                \caption{Flow diagram of the lifespan of a \REVISION{successfully delivered} q-datagram, including inter-layer interactions, from the arrival of the QTCP request up to the reception of the QTCP acknowledgment. Circles represent events, rounded rectangles represent actions, squares represent data structures, and diamonds represent conditional execution. Dashed boxes group elements within the same logical component. Solid arrows determine the flow direction, envelopes and dashed compound arrows indicate data transmission between nodes. The quantum memory management unit \REVISION{introduced in Section \ref{sebsec:switching}} is reported as QMMU, and the \textit{label-slot maps} data structure associates link-level labels with their memory qubit slot. The diagram highlights that the quantum and classical part of a q-datagram are forwarded at slightly different times. While the classical part is forwarded with the classical packet, we refer to the \ac{BSM} as the operation that forwards the quantum part, as the switch frees the memory qubit that held its half of the Bell pair immediately after the \ac{BSM}. \REVISION{In the example, QTCP acknowledgments also report AQM congestion signals, handled within the \textit{Update window size} action. If the quantum memory is full ---this case is not shown here--- the switch drops a q-datagram and sends a congestion notification to the flow source}.
                }
                \label{fig:flow_diagram}
            \end{figure*}
        Now that we have introduced all the architectural elements, we summarize in Fig.~\ref{fig:flow_diagram} all events, actions, and interactions \REVISION{that accompany} the lifespan of a q-datagram. The diagram simplifies Link and Physical layers by only showing the components interacting with the layers above.
        
    \subsection{Considerations}

        Bell pairs are consumable resources for quantum applications, and in principle, both end nodes running an application can serve as flow sources. 
        \ac{QTCP} flows could indeed be \emph{bidirectional}, allowing both nodes to generate requests during the transmission phase. A further modification, with no counterpart in classical networks, would be the possibility of swapping two q-datagrams from the same flow going in opposite directions if they meet at a quantum switch. We call this modification \emph{dynamic rendez-vous}, and while we mention the possibility, we delegate investigating its benefit to future work.



\section{Evaluation}\label{sec:eval}
    
    \subsection{Simulation Setup}
    We \REVISION{simulate} a simple quantum network composed of five quantum nodes connected in a chain. All quantum links, modeled as fibers, are $L=40$km in length, adding up to a total chain length of $160$km. A link controller and an entangled photon source are placed at the midpoint of each link. We consider a fiber attenuation of $0.2$ dB/km and set the joint efficiency of photon emission and frequency coupling at $\eta=0.4$. We do not incorporate the potential losses that may arise when photons are loaded into memories.

    We model \ac{LLEG} as a discrete stochastic process where every sample is an independent, identically distributed geometric random variable $K$ distributed according to $P(K=k) = q(1-q)^{k-1}$, where $k\geq1$ is the number of attempts to success and $q=\eta q_L^2$ is the probability of success of each attempt, where $q_L$ denotes the photon transmission probability from source to switch. The average rate of \ac{LLEG} is thus $\mu=qf_a$ generated pairs per second, where $f_a=10^5$Hz is the attempt frequency of the protocol.
    
    Every memory qubit has a coherence time of $100$ms with a depolarizing noise model. We set the initial quantum state of link-level pairs as Werner states with fidelity $0.99$. Quantum gates and measurements are assumed to be ideal. Every node has $50$ memory qubits per interface and can operate deterministic BSMs \REVISION{on any pair of qubits}. For example, superconducting or ion trap memory technologies \cite{Riebe2008} support this assumption.

    If there are no free memory qubits to accommodate a new link-level pair, the quantum memory management unit drops the oldest buffered q-datagram and uses its memory qubit to buffer \REVISION{a} new pair. In other words, q-datagram queues experience oldest-first drops when they are full.
    
    We assume adjacent nodes share both classical and quantum links so that both types of information can be transmitted along matching topologies. Classical information incurs speed-of-light delay, but is considered reliable.
    
    \subsection{Simulation Results}

    \begin{figure*}
            \begin{centering}
                \subfloat[]{\includegraphics[width=0.49\linewidth]{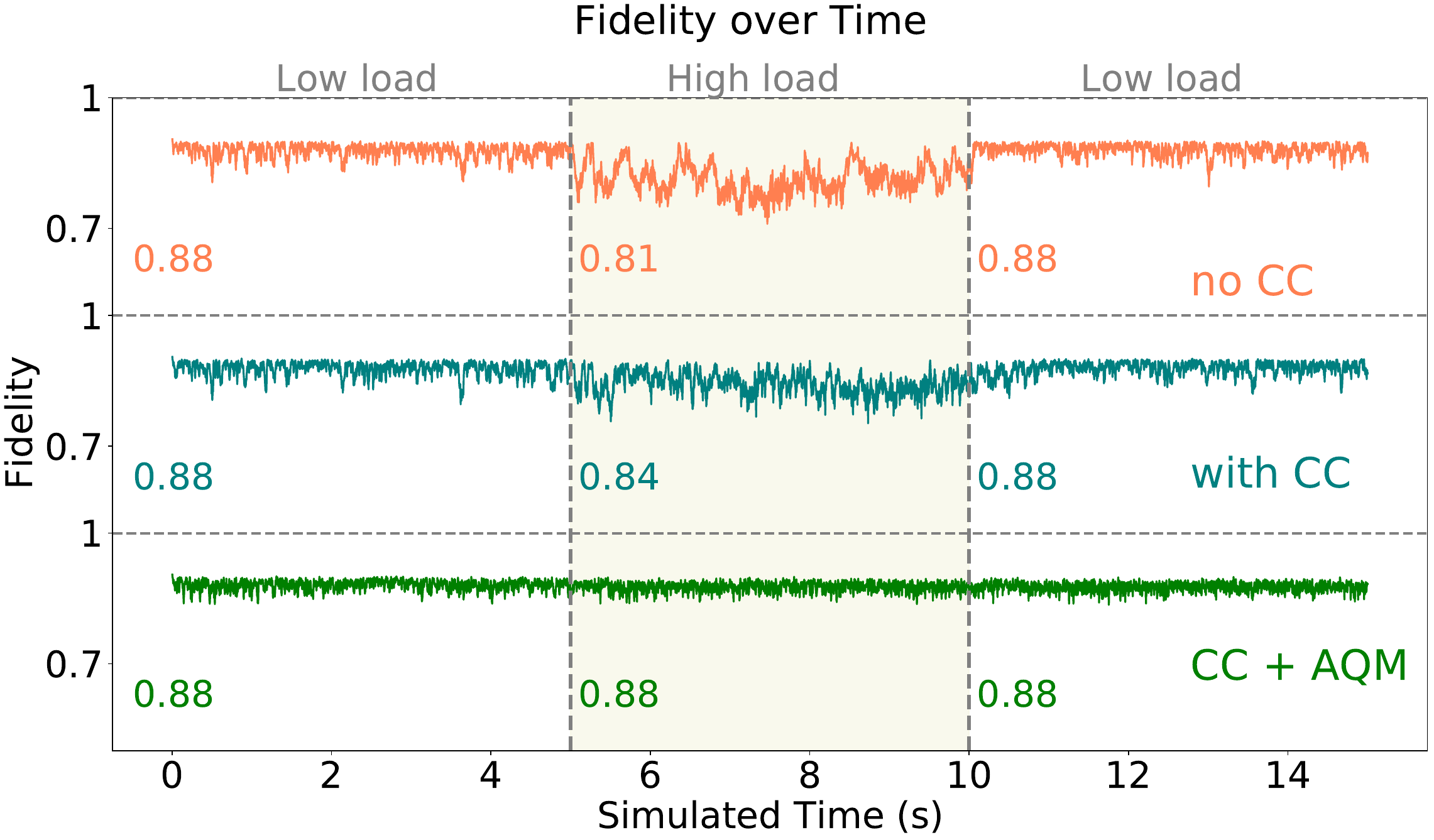} \label{subfig:fid_poisson}} \hfill
                \subfloat[]{\includegraphics[width=0.49\linewidth]{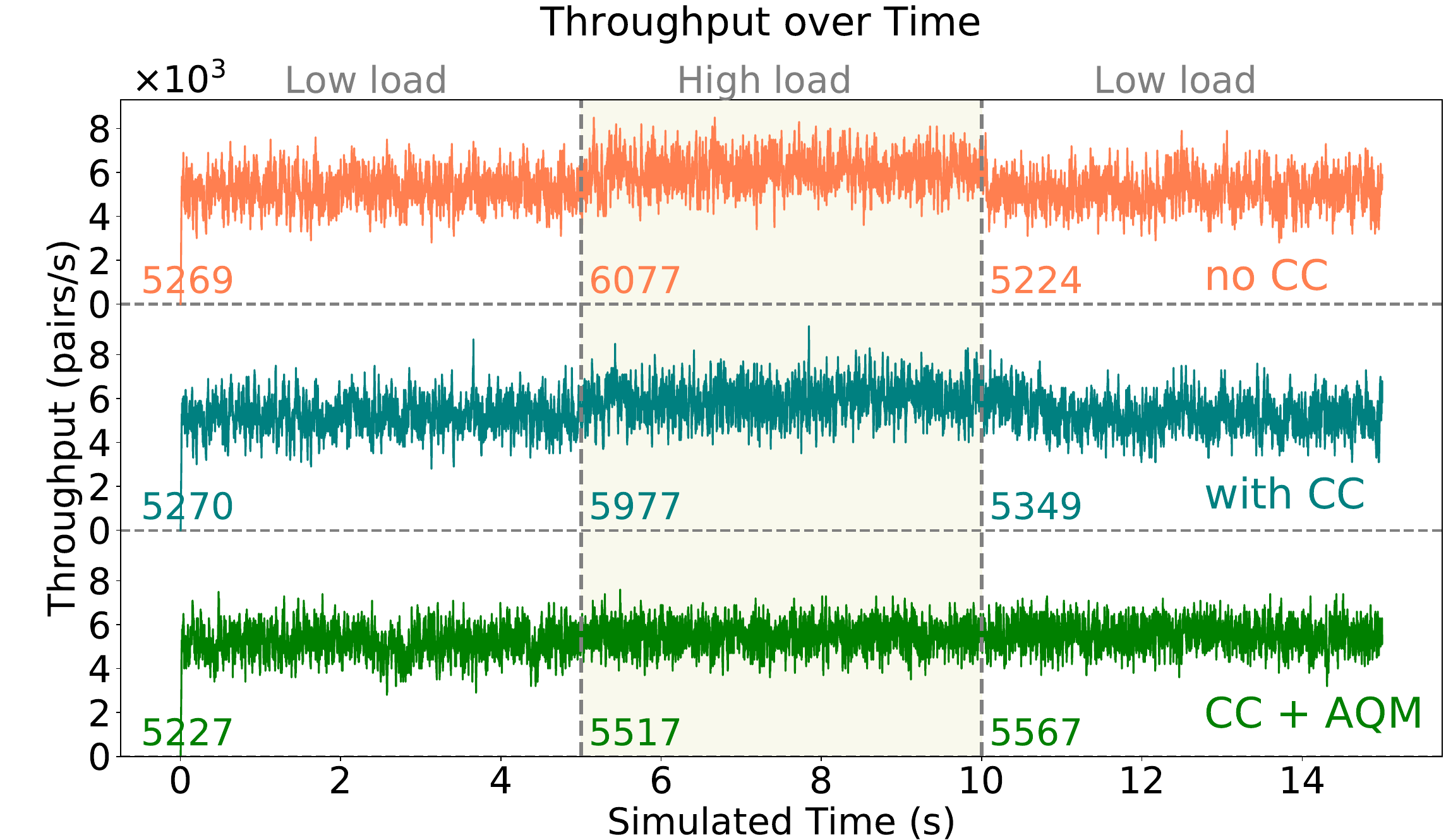}\label{subfig:thr_poisson}}
            \end{centering}
            \centering
            \caption{(a) Fidelity and (b) throughput of delivered \ac{e2e} Bell pairs in a representative simulation of $12$ QTCP flows with Poisson arrivals. The three curves compare the performance of QTCP with no congestion control (orange), with congestion control (teal), and with congestion control plus \ac{AQM} (target buffering time $0.5$ms) (green). ``Low load" regimes (initial and final five seconds) have a load $\rho=0.85$, whereas ``High load" regimes (mid five seconds) have $\rho=0.995$. Throughput is the sliding average on $10$ms. Numbers report the average value across the regime.
            }
            \label{fig:fid_thr}
    \end{figure*}

    \begin{figure*}
            \begin{centering}
                \subfloat[]{\includegraphics[width=0.49\linewidth]{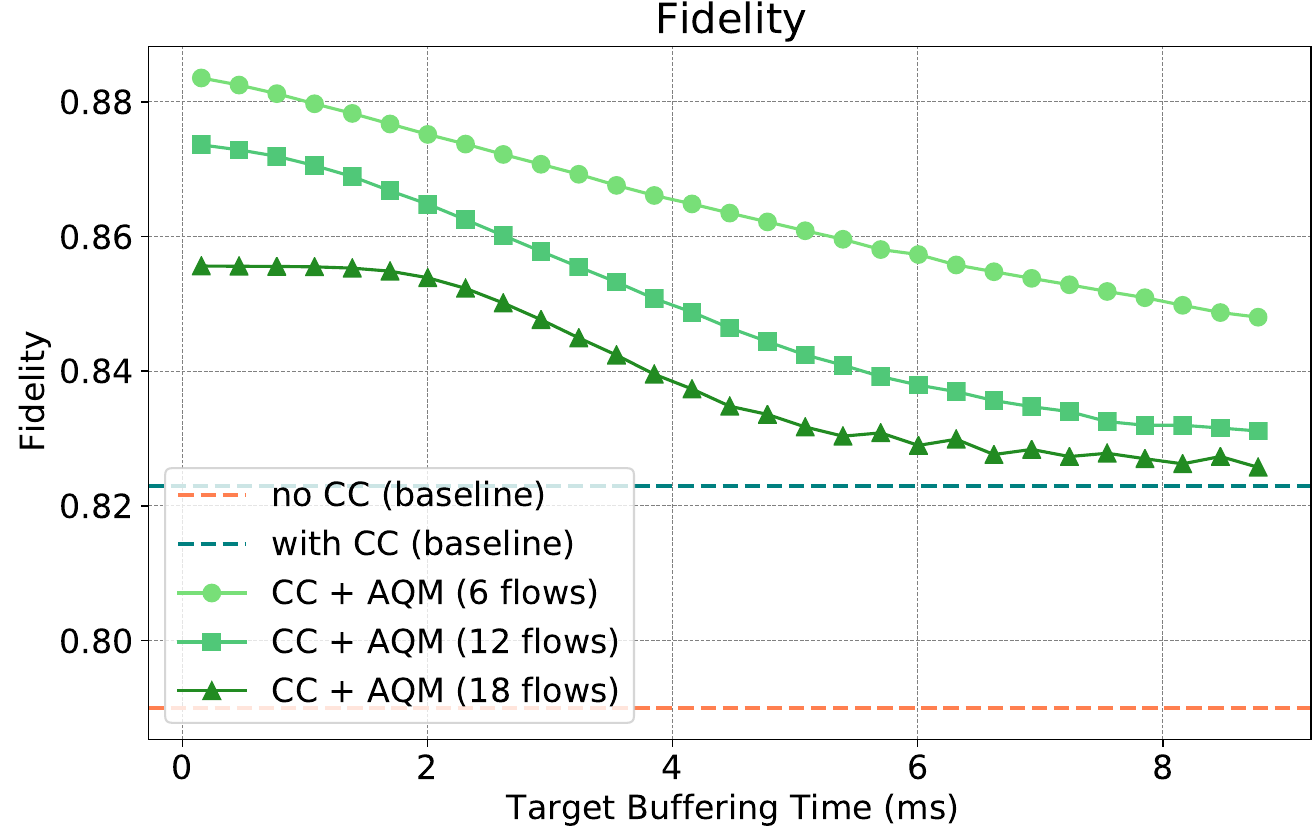}\label{subfig:tune_fid}} \hfill
                \subfloat[]{\includegraphics[width=0.49\linewidth]{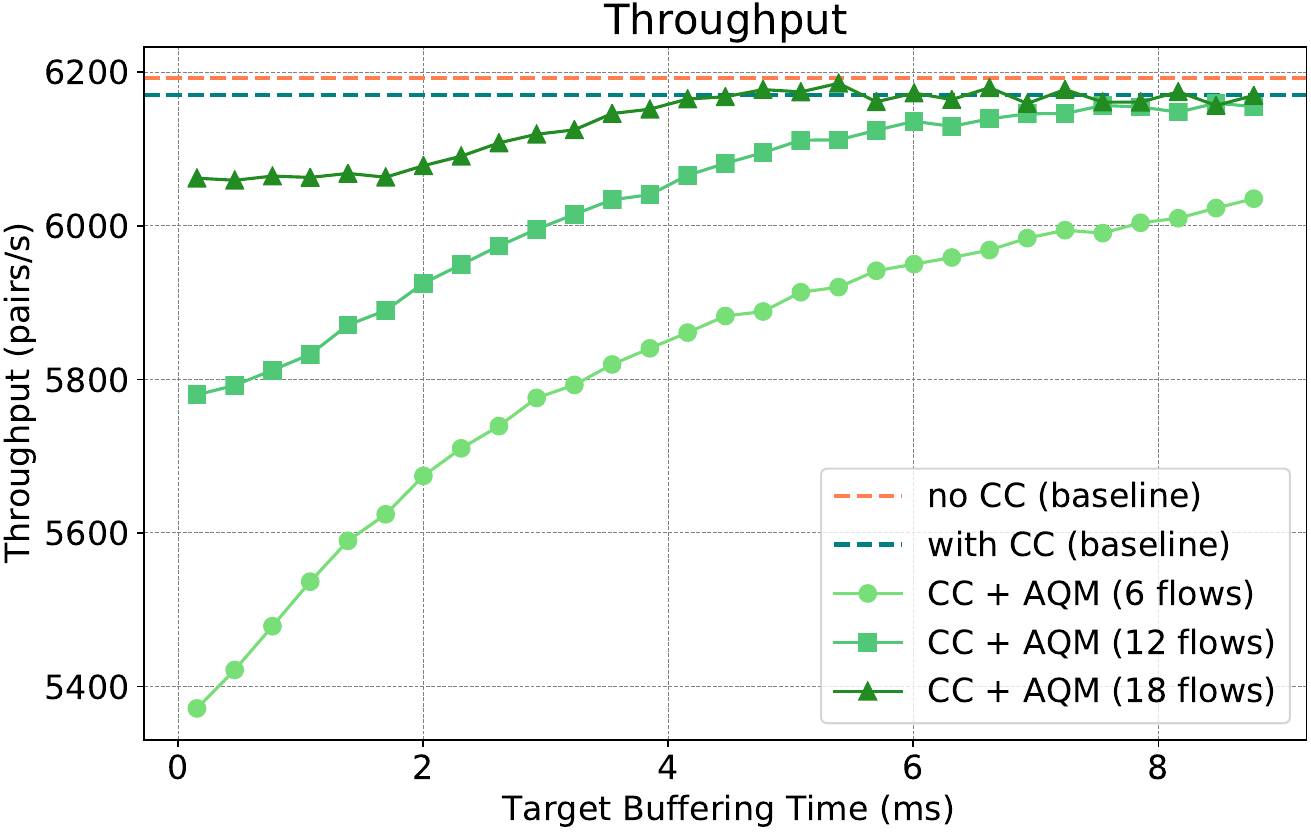}\label{subfig:tune_thr}}
                \\
                \subfloat[]{\includegraphics[width=0.49\linewidth]{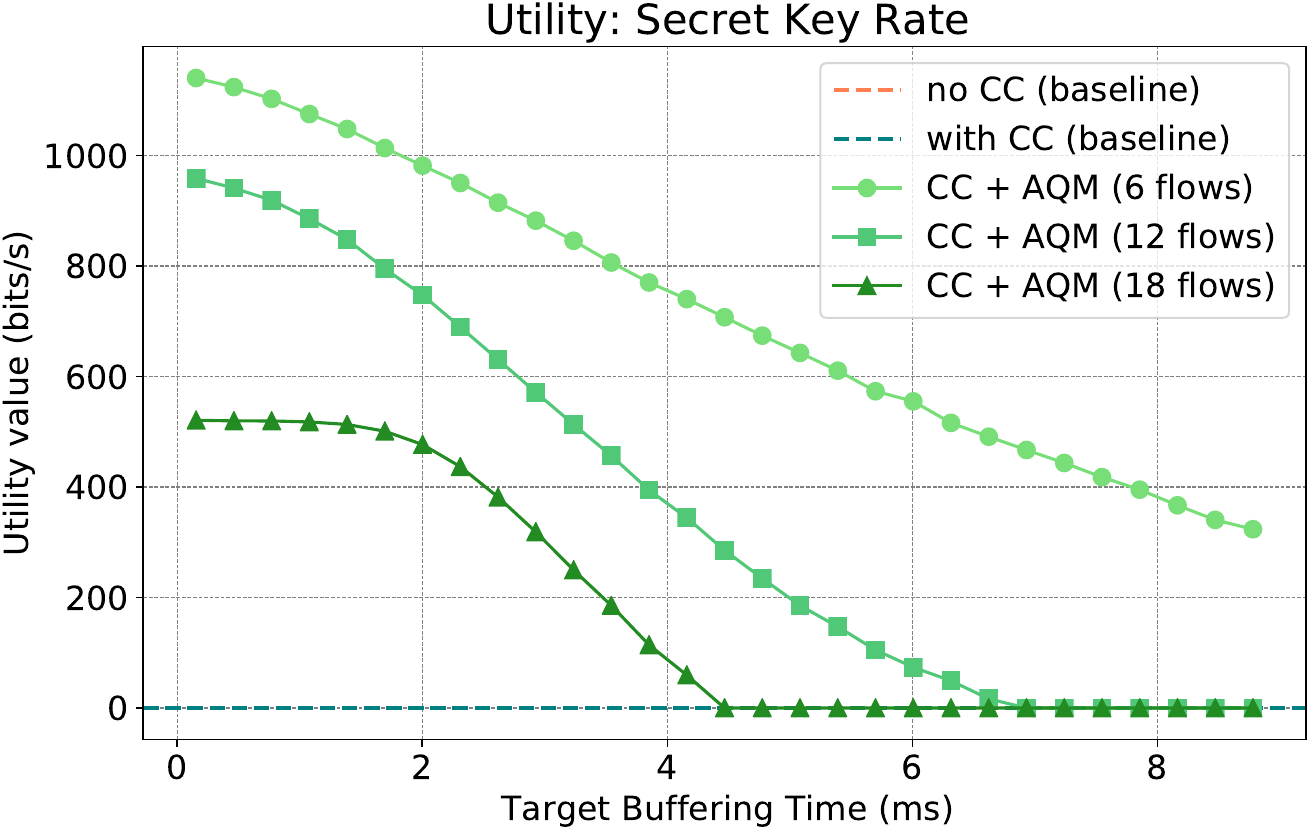}\label{subfig:tune_skr}} \hfill
                \subfloat[]{\includegraphics[width=0.49\linewidth]{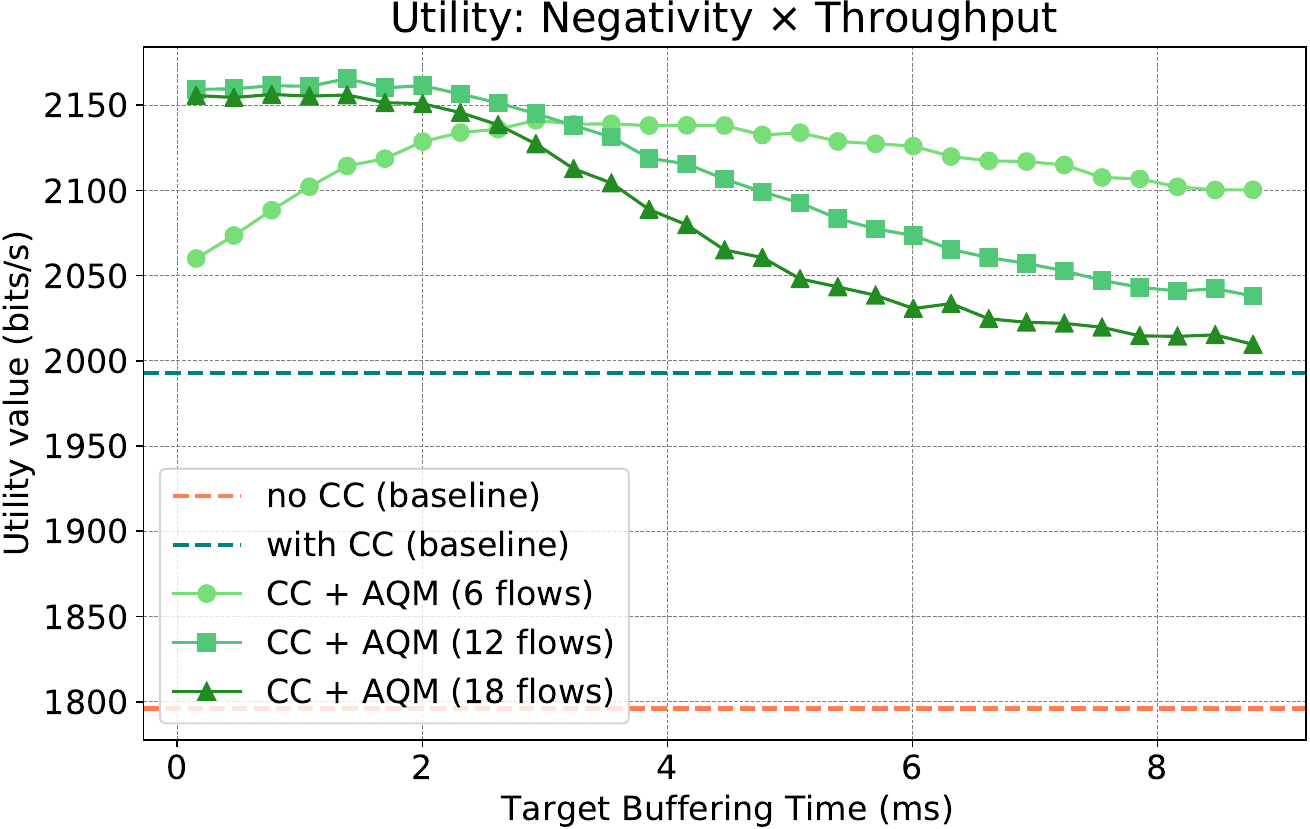}\label{subfig:tune_neg}}
                
            \end{centering}
            \centering
            \caption{Aggregate (a) Fidelity, (b) Throughput, (c) Secret Key Rate of the BB84 protocol, and (d) a utility based on entanglement negativity as a function of the target buffering time. Points are obtained by averaging over the outcome of $16$ independent simulations of $5$ seconds each. Confidence intervals are smaller than or comparable to marker sizes. The two dashed baselines for (a,b,c,d) are obtained by disabling congestion control (teal) and activating AIMD congestion control with no AQM (orange).}
            \label{fig:vs_ref_queue}
    \end{figure*}

    \subsubsection{Proof of Concept}
   First, we allocate $12$ \ac{QTCP} flows between the two end nodes of the chain. Requests arrive according to a Poisson process with an average rate $\lambda/12$ for every flow. The aggregate inter-arrival rate is $\lambda$. We define the load on the network $\rho=\lambda/\mu$.

    Figure~\ref{fig:fid_thr} demonstrates the necessity and benefit of introducing congestion control within \ac{QTCP}. Without congestion control, the average fidelity oscillates depending on the current load, and as the load approaches $1$ (``High Load" regime), roughly five percent of q-datagrams are lost, causing significant resource waste. Adding \ac{AIMD} congestion control prevents most of this resource wastage and improves fidelity. However, fidelity is further improved through the introduction of AQM because flows react to congestion before memory overflow (and\REVISION{, therefore,} loss) occurs.  Summarizing the results, \ac{AQM} (i) reduces losses, (ii) improves average fidelity, (iii) decreases fidelity variance, and (iv) makes average fidelity insensitive to load variations.
    
    Fig.~\ref{subfig:thr_poisson} shows that the advantages of applying \ac{AQM} come at about ten percent cost in terms of throughput in the ``High Load" regime. AQM causes requests waiting for admission to accumulate for the \REVISION{entire} duration of the ``High Load" regime. This admission queue then progressively empties during the following ``Low Load" regime. That explains why throughput does not decrease when the load drops.

    \subsubsection{Exposing the Right Knobs} 
   We now evaluate the impact of changing the target buffering time of \ac{PI} controllers. To this aim, we run a second set of simulations with $6$, $12$, and $18$ QTCP flows. We switch to a scenario with infinite demands, meaning that QTCP flows always have requests ready. In the first two plots of Fig. \ref{fig:vs_ref_queue}, we show average fidelity and throughput as functions of the target buffering time. Plots show that these two metrics conflict with each other and appropriate tuning will depend on application requirements.  In order to illustrate this, we associate a utility, function of fidelity and throughput, to an application. Fig. \ref{fig:vs_ref_queue} illustrates the effect of target buffering time on two different utility functions: Secret Key Rate of BB84 and an entanglement negativity-based utility, drawn from \cite{Vardoyan_2023}. The best tuning differs for the two cases, as the maximum value corresponds to the shortest buffering time ($0.15$ms) for Secret Key Rate (Fig. \ref{subfig:tune_skr}), and to an intermediate value (different for every curve) in the case of the negativity-based utility (Fig. \ref{subfig:tune_neg}).
    
    Finally, we observe that the number of flows passing through the network has some impact on performance, as more flows make it harder for \ac{PI} controllers to manage queues. This is especially true since we are working with queue sizes much smaller than those in classical networks and increased variability in service times (due to \ac{LLEG}). One option  to mitigate this phenomenon --left for future investigation-- is to use adaptive tuning of the \ac{PI} controller parameters.

\section{Conclusion}\label{sec:conclusion}
An approach to building the Quantum Internet is presented in this paper by abstracting quantum-specific elements and adopting a packet-switched network architecture. We leveraged well-established classical Internet algorithms and protocols for  congestion control and active queue management. Specifically, we showcased how to tailor these classical methods to address unique quantum networking problems such as memory decoherence. Our simulation results validated  this approach by showing that the proposed architecture maintains high utilization, while ensuring entanglement fidelity around a target value.

The implications of this work are significant, as it opens up new avenues for quantum network design by demonstrating that classical networking principles can be effectively applied to quantum systems. This reframing of quantum networks simplifies the design process and allows the integration of existing technologies, potentially accelerating the development of a fully operational Quantum Internet. Future research can build on this foundation, exploring combinations of classical and quantum-native tools.

In summary, this paper lays the groundwork for a practical and scalable quantum network infrastructure, bridging part of the gap between classical and quantum networking paradigms.

\section*{Acknowledgment}
This research was supported in part by the NSF
grant CNS-1955744, NSF- ERC Center for Quantum Networks grant
EEC-1941583, and DOE  Grant AK0000000018297.
\REVISION{This research is sponsored in part by PiQSci projects of Advanced Scientific Computing Research program, U.S. Department of Energy. This manuscript has been co-authored by UT-Battelle, LLC, under contract DE-AC05-00OR22725 with the US Department of Energy (DOE). The US government retains and the publisher, by accepting the article for publication, acknowledges that the US government retains a nonexclusive, paid-up, irrevocable, worldwide license to publish or reproduce the published form of this manuscript, or allow others to do so, for US government purposes. DOE will provide public access to these results of federally sponsored research in accordance with the DOE Public Access Plan (http://energy.gov/downloads/doe-public-access-plan)}


\bibliographystyle{ieeetr}
\bibliography{refs,additional-refs}

\end{document}